\documentclass{PoS}

\title{EOS in 2+1 flavor QCD with improved Wilson quarks by the fixed-scale approach}

\ShortTitle{EOS in 2+1 flavor QCD with improved Wilson quarks}

\author{\speaker{T. Umeda}\\
        Graduate School of Education, Hiroshima University, Hiroshima 739-8524, Japan\\
E-mail: \email{tumeda@hiroshima-u.ac.jp}}

\author{S. Aoki, K. Kanaya, H. Ohno\\
        Graduate School of Pure and Applied Sciences, University of Tsukuba, Tsukuba, Ibaraki 305-8571, Japan}

\author{S. Ejiri\\
        Graduate School of Science and Technology, Niigata University, Niigata 950-2181, Japan}

\author{T. Hatsuda \\
        Department of Physics, The University of Tokyo, Tokyo 113-0033, Japan}

\author{Y. Maezawa\\
        Mathematical Physics Laboratory, RIKEN Nishina Center, Saitama 351-0198, Japan }

\author{(WHOT-QCD Collaboration)}

\abstract{We present the status of our study on the equation of state in 2+1 flavor QCD with non-perturbatively improved Wilson quarks coupled with the RG improved glue.  
We apply the $T$-integration method to  non-perturbatively calculate the equation of state by the fixed-scale approach.}

\FullConference{The XXVIII International Symposium on Lattice Field Theory, Lattice2010\\
		June 14-19, 2010\\
		Villasimius, Italy}

\begin{document}

\section{Introduction}
Lattice QCD simulations enable us to calculate QCD thermodynamic quantities, 
such as the equation of state (EOS) in quark gluon plasma (QGP),
non-perturbatively.
The lattice results serve indispensable elements in understanding the
nature of QGP in heavy Ion collision experiments, 
e.g.\  as inputs to the hydrodynamical space-time evolution of QGP.

In the last decade, importance of dynamical quarks in EOS became gradually apparent.
However study of EOS with dynamical quarks requires one of the most expensive 
calculations in lattice QCD.
It is sometimes hard to control lattice artifacts in the EOS calculations 
even with recent high performance computers.

First of all, the lattice should be fine enough to keep lattice artifacts small.
In the conventional fixed $N_t$ approach ($N_t$ is the temporal lattice extent), however,
the lattice spacing $a$ is large for low temperatures.
Such coarse lattices not only cause lattice artifacts in the observables
but also introduce uncertainties in the determination of the line of constant physics (LCP).
Because the integral method \cite{inte_method} requires an integration from a low $T$,
the uncertainties in the low $T$ region affect EOS in the whole range of $T$.
Therefore, it is important to keep the lattice fine in the whole range of $T$.
On the other hand, in the high $T$ region, small values of $a$ there lead to small lattice volumes which may cause finite volume effects.

In this study, we adopt the fixed-scale approach, in which we vary $T$ by 
changing $N_t$ at a fixed $a$, by fixing all coupling parameters \cite{Tintegral,Kanaya:2009nf}. 
In this approach 
the problems mentioned above are in part resolved. 
Among others, we do not need to determine LCP at all:
Since the coupling parameters are common to all temperatures,
the condition to follow a LCP is automatically satisfied.
Furthermore, when we borrow zero-temperature configurations from a large scale spectrum study, 
the lattice spacings are in the scaling region and are  
smaller than those used in conventional fixed $N_t$ studies around the 
transition temperature $T_c$. 
In the high $T$ region, on the other hand, while the lattice volume is kept fixed, we suffer from lattice artifacts due to small $N_t$.
Therefore, the fixed-scale approach is complemental to the fixed $N_t$ approach.
A good feature of the fixed-scale approach is that the computational cost for zero-temperature simulations, which is a big burden in the fixed $N_t$ approach, can be largely reduced.

Secondly, the choice of lattice action has important implications for the lattice artifacts.
Although recent large scale simulations of finite temperature/density QCD mostly adopt 
staggered-type lattice quarks \cite{Borsanyi:2010cj,Bazavov:2010sb}, 
a rigorous proof is missing for the staggered-type quarks about the existence of the continuum limit with desired number of quark flavors.
We thus adopt Wilson-type quarks with which the continuum limit is guaranteed for any number of flavors.
To remove $O(a)$ lattice artifacts, we adopt a clover-improved Wilson quark action \cite{Sheikholeslami:1985ij} with the clover coefficient $c_{SW}$ non-perturbatively evaluated by the Schr\"{o}dinger functional method \cite{Aoki:2005et}.
Note that the Schr\"{o}dinger functional method also requires fine lattices.

Thirdly, EOS is sensitive to quark masses and to the number of flavors.
Therefore, for a realistic calculation, we should incorporate the strange quark and try to extrapolate to the physical quark mass point.
Even an effect of the charm quark on EOS has been recently discussed \cite{Cheng:2007wu}.
With Wilson-type quarks, however, systematic calculation of EOS has been limited to the case of two-flavor QCD with rather heavy quarks \cite{CP-PACS-EOS}.

We extend the study to the case of 2+1 flavor QCD.
The computational cost can be in part reduced by adopting the fixed-scale approach
by borrowing zero-temperature configurations from large-scale spectrum studies. 
We adopt configurations of 2+1 flavor 
QCD with non-perturbatively improved Wilson quarks, generated by the CP-PACS+JLQCD 
Collaboration \cite{Aoki:2005et,Ishikawa:2007nn}.
The configurations are available on the ILDG.
Using the same coupling parameters, we carry out finite temperature simulations varying $N_t$.
Details of the simulation parameters are discussed in Sect.\ref{sec:setup}.

\section{$T$-integration method}

In ref.\cite{Tintegral}, we have developed the ``$T$-integration method'' to evaluate 
the pressure non-perturbatively in the fixed-scale approach:
Using a thermodynamic relation valid at vanishing chemical potential
\begin{eqnarray}
T \frac{\partial}{\partial T} \left( \frac{p}{T^4} \right) =
\frac{\epsilon-3p}{T^4},
\end{eqnarray}
we obtain
\begin{eqnarray}
\frac{p}{T^4} = \int^{T}_{T_0} dT \, \frac{\epsilon - 3p}{T^5}
\label{eq:Tintegral}
\end{eqnarray}
with $p(T_0) \approx 0$.
Here, the trace anomaly $\epsilon -3p$ is calculated as usual at each temperature.
In the fixed-scale approach, $T$ is restricted to discrete values 
due to the discreteness of $N_t$.
For the integration of (\ref{eq:Tintegral}), we need to interpolate 
the data with respect to $T$. 
The systematic error from the interpolation should be checked.
Note that, because the scale is common for all data points in the 
fixed scale approach, $T$ is determined without errors besides the 
common overall factor $1/a$.

\section{Lattice setup}
\label{sec:setup}
As the zero-temperature configurations, we adopt the results of a 2+1 flavor QCD spectrum study with 
improved Wilson quarks by the CP-PACS+JLQCD Collaboration \cite{Ishikawa:2007nn}.
The QCD action $S=S_g+S_q$ is defined by the
RG-improved gauge action $S_g$ and the clover-improved Wilson quark action
$S_q$,
\begin{eqnarray}
S_g &=& -\beta\left\{ 
\sum_{x,\mu>\nu}c_0W^{1\times 1}_{\mu\nu}(x)
+\sum_{x,\mu,\nu}c_1W^{1\times 2}_{\mu\nu}(x)
\right\},\\
S_q &=& \sum_{f=u,d,s}\sum_{x,y} \bar{q}_x^f D_{x,y}q_y^f, \\
D_{x,y} &=& \delta_{x,y}-\kappa_f
\sum_\mu\{ (1-\gamma_\mu)U_{x,\mu}\delta_{x+\hat{\mu},y}
+(1+\gamma_\mu)U^\dagger_{x-\hat{\mu},\mu}\delta_{x-\hat{\mu},y}
\}\nonumber\\
&&-\delta_{x,y}c_{SW}\kappa_f\sum_{\mu>\nu}\sigma_{\mu\nu}
F_{\mu\nu},
\end{eqnarray}
where $c_{SW}$ is non-perturbatively determined as a function of $\beta$ \cite{Aoki:2005et}.
Among the simulation points by the CP-PACS+JLQCD Collaboration, we choose $\beta=2.05$, $\kappa_{ud}=0.1356$ and 
$\kappa_s=0.1351$ which correspond to the smallest lattice spacing
$a\simeq 0.07$fm, and the lightest $u$ and $d$ quark masses 
$m_\pi/m_\rho\simeq0.63$ and $m_{\eta_{ss}}/m_{\phi}\simeq0.74$ in the study.
The lattice size is $28^3 \times 56$ and the statistics is about
6000 trajectories.

\begin{figure}[bt]
  \begin{center}
    \begin{tabular}{ccc}
    \includegraphics[width=75mm]{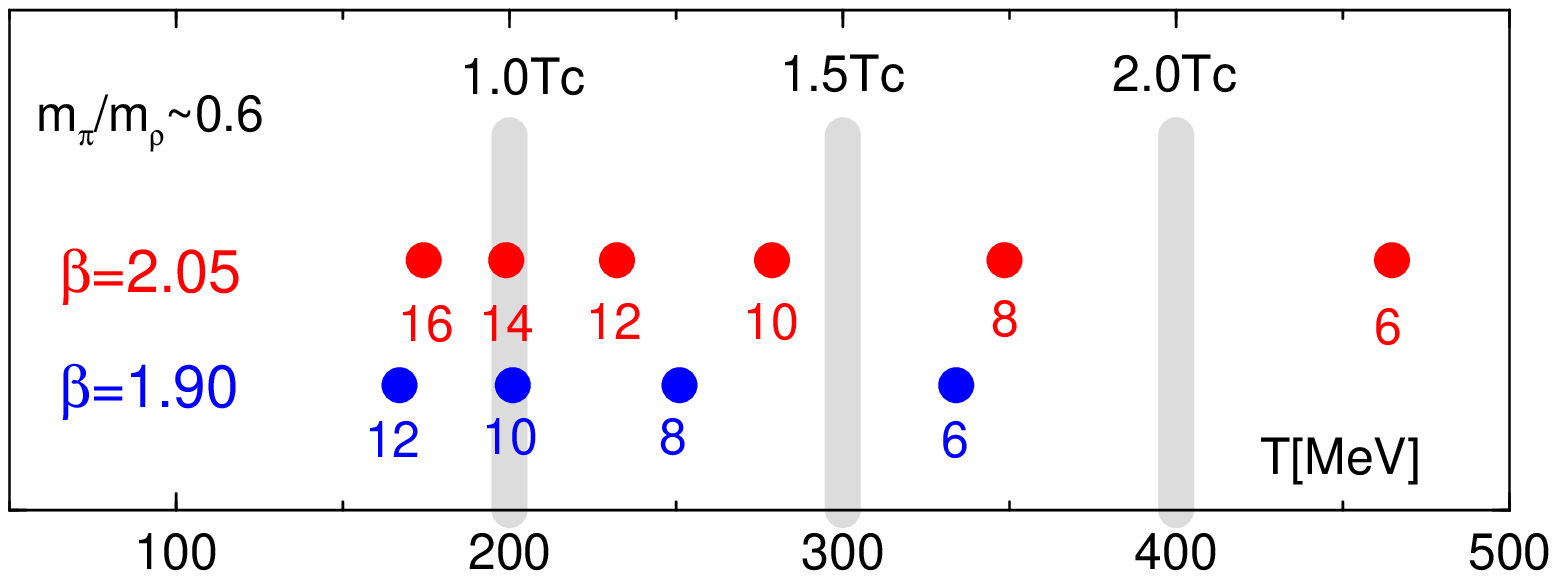} & &
    \includegraphics[width=65mm]{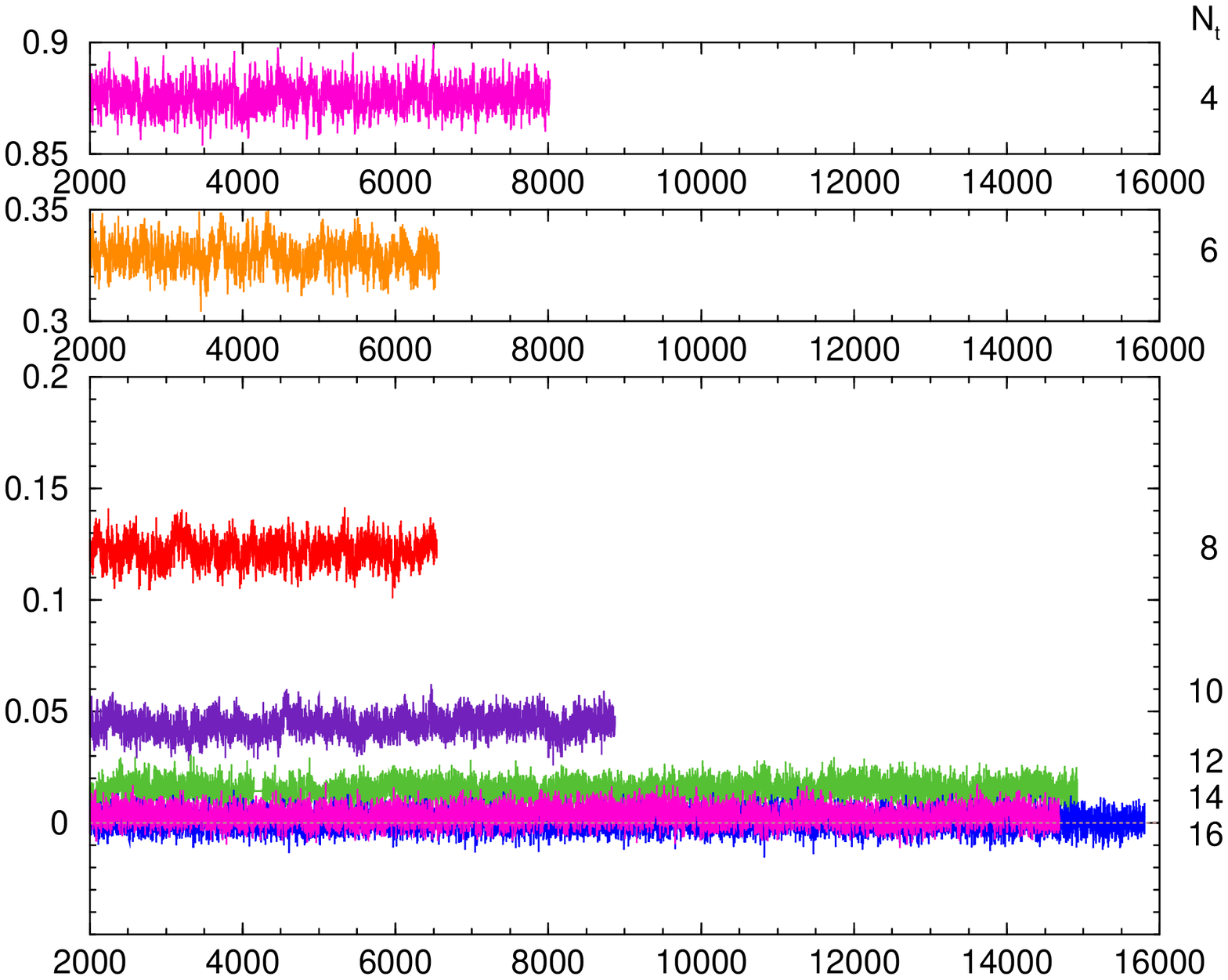}
    \end{tabular}
    \caption{
    {\em Left:} Estimated temperatures of our lattice setup for each 
    $N_t$ at $\beta=2.05$ and 1.90. Physical scale is determined by 
    Sommer scale $r_0=0.5$fm. 
    {\em Right:} The Polyakov loop time history as a status of finite
    temperature $N_f=2+1$ QCD simulation with improved Wilson quarks. 
    }
    \label{fig1}
  \end{center}
\end{figure}

Using the same coupling parameters as the zero-temperature simulation, 
we are generating finite temperature configurations on 
$32^3\times N_t$ lattices with $N_t=4$, 6, $\cdots$, 16.
Setting the lattice scale by $r_0=0.5$fm, the 
temperatures for these $N_t$ at $\beta=2.05$ are shown  
in the left panel of Fig.~\ref{fig1}.
The pseudo-critical temperature is expected to be $N_t \sim 14$. 
Current status of our finite temperature simulation is shown in the right 
panel of Fig.\ref{fig1}.
Here $1$ trajectory is equal to $0.5$ molecular dynamics step.

The trace anomaly $(\epsilon-3p)/T^4$ for our action is given by
\begin{eqnarray}
\frac{\epsilon-3p}{T^4}&=&=\frac{N_t^3}{N_s^3}
\left(
{a\frac{\partial\beta}{\partial a}}
\left\langle
\frac{\partial S}{\partial\beta}
\right\rangle_{sub}
+{a\frac{\partial\kappa_{ud}}{\partial a}}
\left\langle
\frac{\partial S}{\partial \kappa_{ud}}
\right\rangle_{sub}
+{a\frac{\partial\kappa_s}{\partial a}}
\left\langle
\frac{\partial S}{\partial \kappa_s}
\right\rangle_{sub}
\right) \label{eq:tranom}\\
\left\langle 
\frac{\partial S}{\partial \beta} 
\right\rangle &=&N_s^3N_t\left(
-\left\langle
\sum_{x,\mu>\nu}c_0W^{1\times 1}_{\mu\nu}(x)
+\sum_{x,\mu,\nu}c_1W^{1\times 2}_{\mu\nu}(x)
\right\rangle \right.\nonumber\\
&&\left.
+N_f\frac{\partial c_{SW}}{\partial \beta}\kappa_f
\left\langle
\sum_{x,\mu>\nu}\mbox{Tr}^{(c,s)}\sigma_{\mu\nu}F_{\mu\nu}
(D^{-1})_{x,x}
\right\rangle
\right)\label{eq:dsdb}\\ 
\left\langle 
\frac{\partial S}{\partial \kappa_f} 
\right\rangle &=&N_fN_s^3N_t\left(
\left\langle
\sum_{x,\mu}\mbox{Tr}^{(c,s)}
\{(1-\gamma_\mu)U_{x,\mu}(D^{-1})_{x+\hat{\mu},x}
+(1+\gamma_\mu)U^\dagger_{x-\hat{\mu},\mu}
(D^{-1})_{x-\hat{\mu},x}
\}
\right\rangle \right.\nonumber\\
&&\left.
+c_{SW}
\left\langle
\sum_{x,\mu>\nu}\mbox{Tr}^{(c,s)}\sigma_{\mu\nu}F_{\mu\nu}
(D^{-1})_{x,x}
\right\rangle
\right)\label{eq:dsdk}
\end{eqnarray}
where $\langle\cdots\rangle_{sub}$ means that the $T=0$ value is subtracted.
To evaluate the traces in (\ref{eq:dsdb}) and (\ref{eq:dsdk})
we apply a random noise method with complex U(1) random numbers
\cite{Ejiri:2009hq}.
The number of noise for each of the color and spinor indices is 1.

\section{Beta functions}
\label{sect:beta}
The beta functions are required to calculate the trace anomaly (\ref{eq:tranom}).
The beta functions are obtained through the coupling parameter dependence
of zero-temperature observables on a LCP.
In our previous report \cite{Kanaya:2009nf}, we have tried to evaluate the beta functions by an inverse matrix method \cite{CP-PACS-EOS}.
It turned out that, although the results for EOS are consistent with an expectation from the two-flavor case,  errors in the beta functions for the hopping parameters are large. 
To resolve the problem, we adopt an alternative method, the direct fit method \cite{CP-PACS-EOS}, 
to estimate the beta functions.
We use the results for the hadron spectrum by the CP-PACS+JLQCD Collaboration \cite{Ishikawa:2007nn}
at three $\beta$'s, five $\kappa_{ud}$'s and two $\kappa_s$'s, 
i.e. totally 30 simulation points at $T=0$. 

\begin{figure}[bt]
  \begin{center}
    \begin{tabular}{ccc}
    \includegraphics[width=46mm]{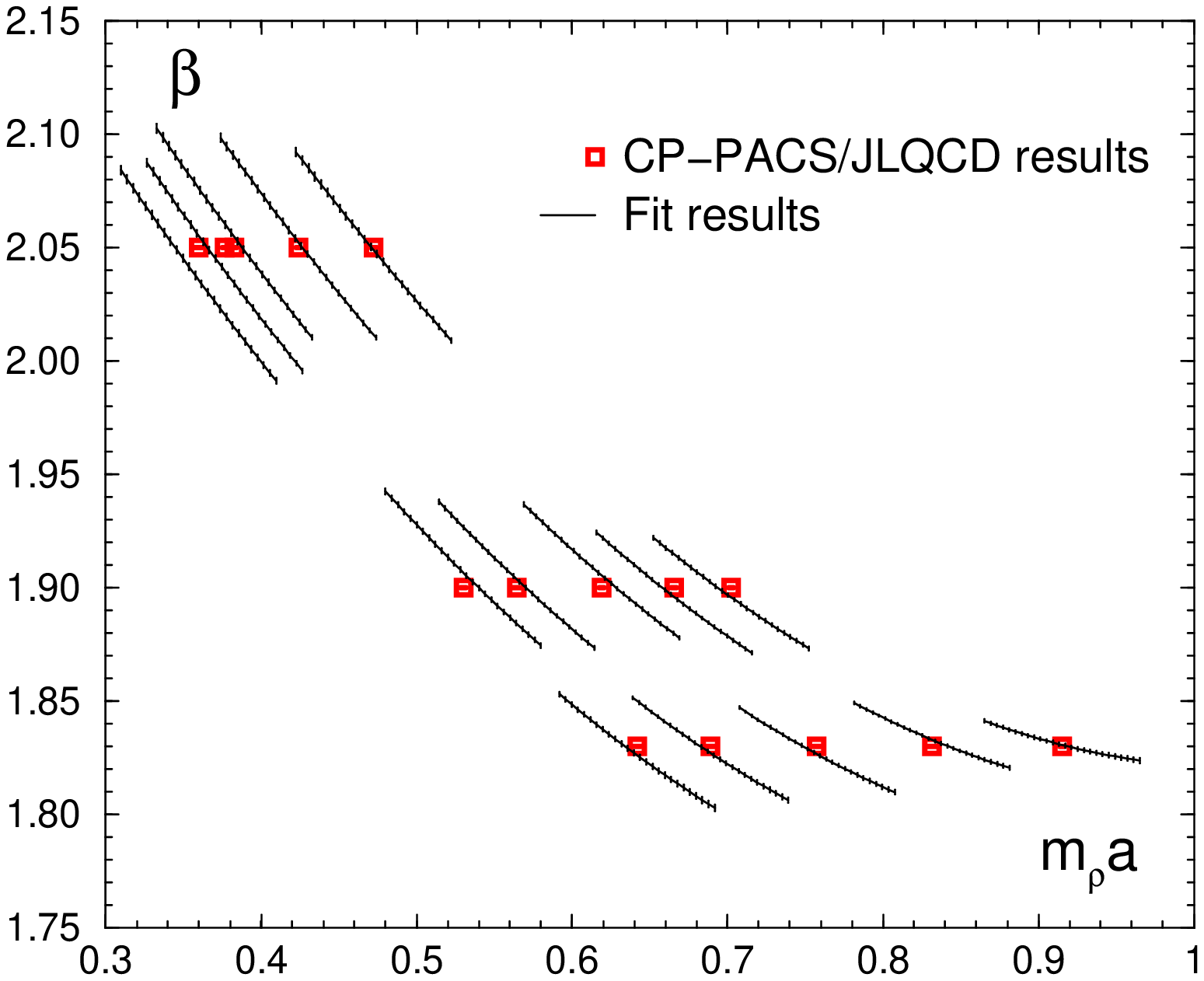} & 
    \includegraphics[width=46mm]{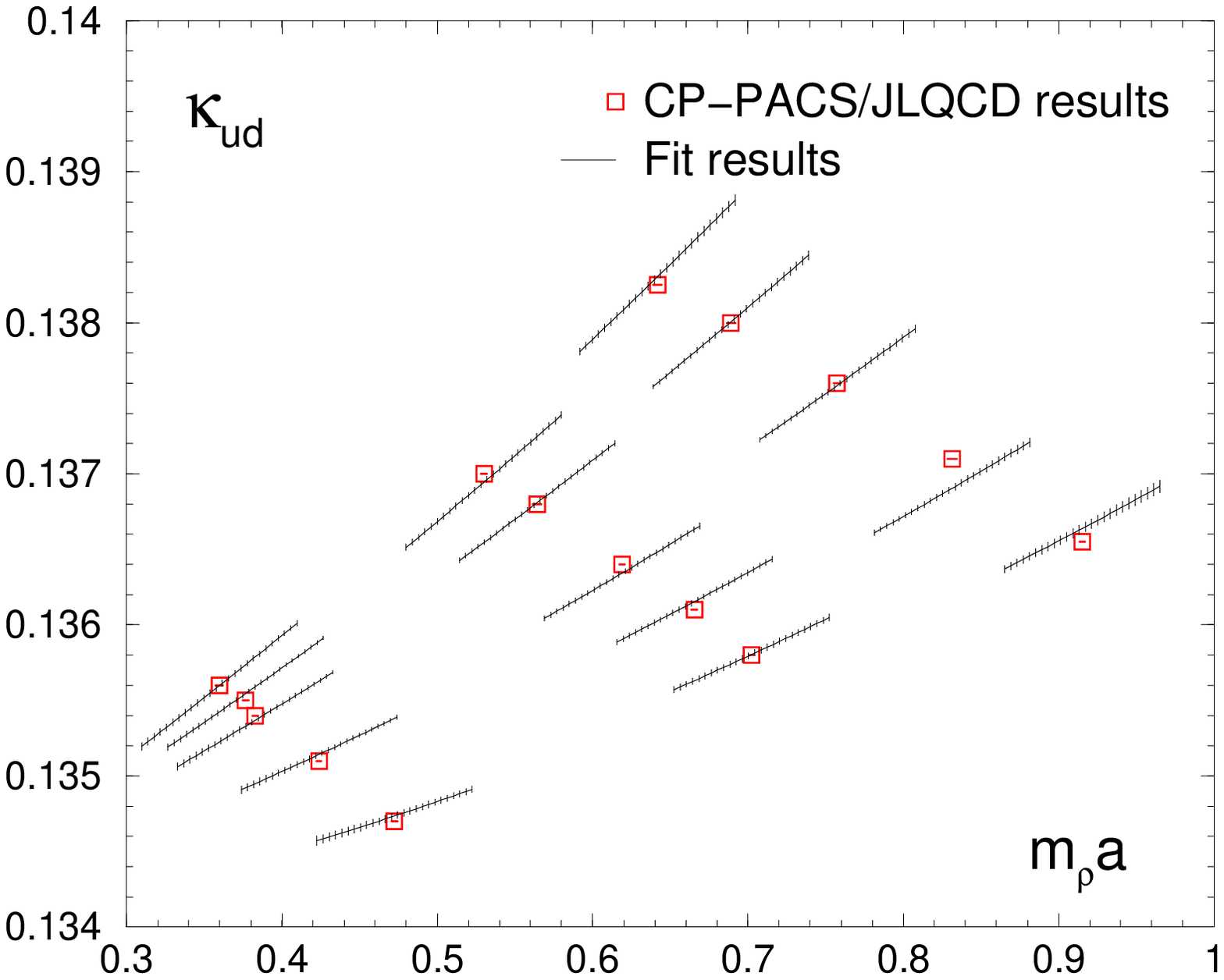} &
    \includegraphics[width=46mm]{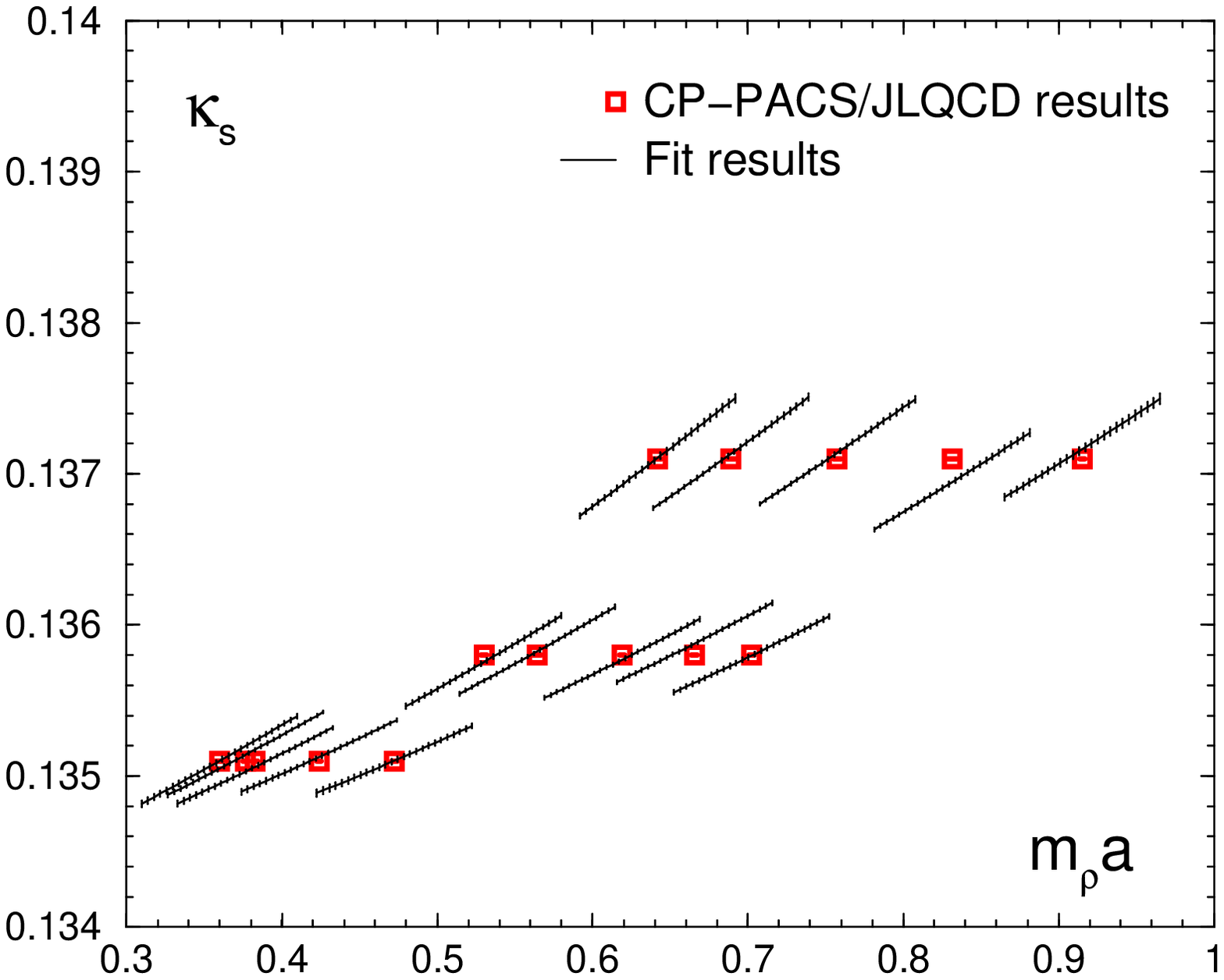}
    \end{tabular}
    \caption{
    The global fit for coupling parameters, {\em Left:} $\beta$,
    {\em Center:} $\kappa_{ud}$, {\em Left:} $\kappa_s$, 
    as the function of $m_\rho a$. Square symbols show coupling 
    parameters in CP-PACS/JLQCD study. The Solid lines show the global
    fit results for each simulation point with corresponding 
    $m_\rho/m_\pi$ and $m_{\eta_{ss}}/m_\phi$. 
    }
    \label{fig2}
  \end{center}
\end{figure}

We fit the coupling parameters, $\beta$, $\kappa_{ud}$ and $\kappa_s$ 
as a polynomial function of three observables $am_\rho$, $m_\pi/m_\rho$ and $m_{\eta_{ss}}/m_\phi$ up to the 2nd order, 
\begin{eqnarray}
\left(
\begin{array}{c}
\beta \\
\kappa_{ud} \\
\kappa_s
\end{array}
\right)
&=& \vec{c_0} + \vec{c_1}(am_\rho) + \vec{c_2}(am_\rho)^2
+ \vec{c_3}\left(\frac{m_\pi}{m_\rho}\right) 
+ \vec{c_4}\left(\frac{m_\pi}{m_\rho}\right)^2
+ \vec{c_5}(am_\rho)\left(\frac{m_\pi}{m_\rho}\right) \nonumber\\
&&+ \vec{c_6}\left(\frac{m_{\eta_{ss}}}{m_\phi}\right) 
+ \vec{c_7}\left(\frac{m_{\eta_{ss}}}{m_\phi}\right)^2
+ \vec{c_8}(am_\rho)\left(\frac{m_{\eta_{ss}}}{m_\phi}\right) 
+ \vec{c_9}\left(\frac{m_\pi}{m_\rho}\right) 
\left(\frac{m_{\eta_{ss}}}{m_\phi}\right).
\label{eq:beta}
\end{eqnarray}
Figure \ref{fig2} shows the results of the global fits (\ref{eq:beta}) as functions of $m_\rho a$.  
The fits lead to reasonable $\chi^2/$dof $\sim 1$ 
except for the $\beta$ fit whose $\chi^2/$dof is about 5.

We define LCP by fixing 
$m_\pi/m_\rho$ and $m_{\eta_{ss}}/m_\phi$.
Therefore, in (\ref{eq:beta}), the lattice spacing dependence on a LCP appears through the terms containing $am_\rho$.
The beta functions are thus calculated from the coefficients
$\vec{c_1}$, $\vec{c_2}$, $\vec{c_5}$ and $\vec{c_8}$.
From the fits we obtain preliminary values for the beta functions: 
${\displaystyle a\frac{\partial \beta}{\partial a} = -0.334(4)}$, 
${\displaystyle a\frac{\partial \kappa_{ud}}{\partial a} = 0.00289(6)}$ and 
${\displaystyle a\frac{\partial \kappa_s}{\partial a} = 0.00203(5)}$ 
at our simulation point.
Here the errors are statistical only.
Estimation of systematic errors is left for future investigations.

\section{Equation of state}
\begin{figure}[bt]
  \begin{center}
    \begin{tabular}{cc}
    \includegraphics[width=70mm]{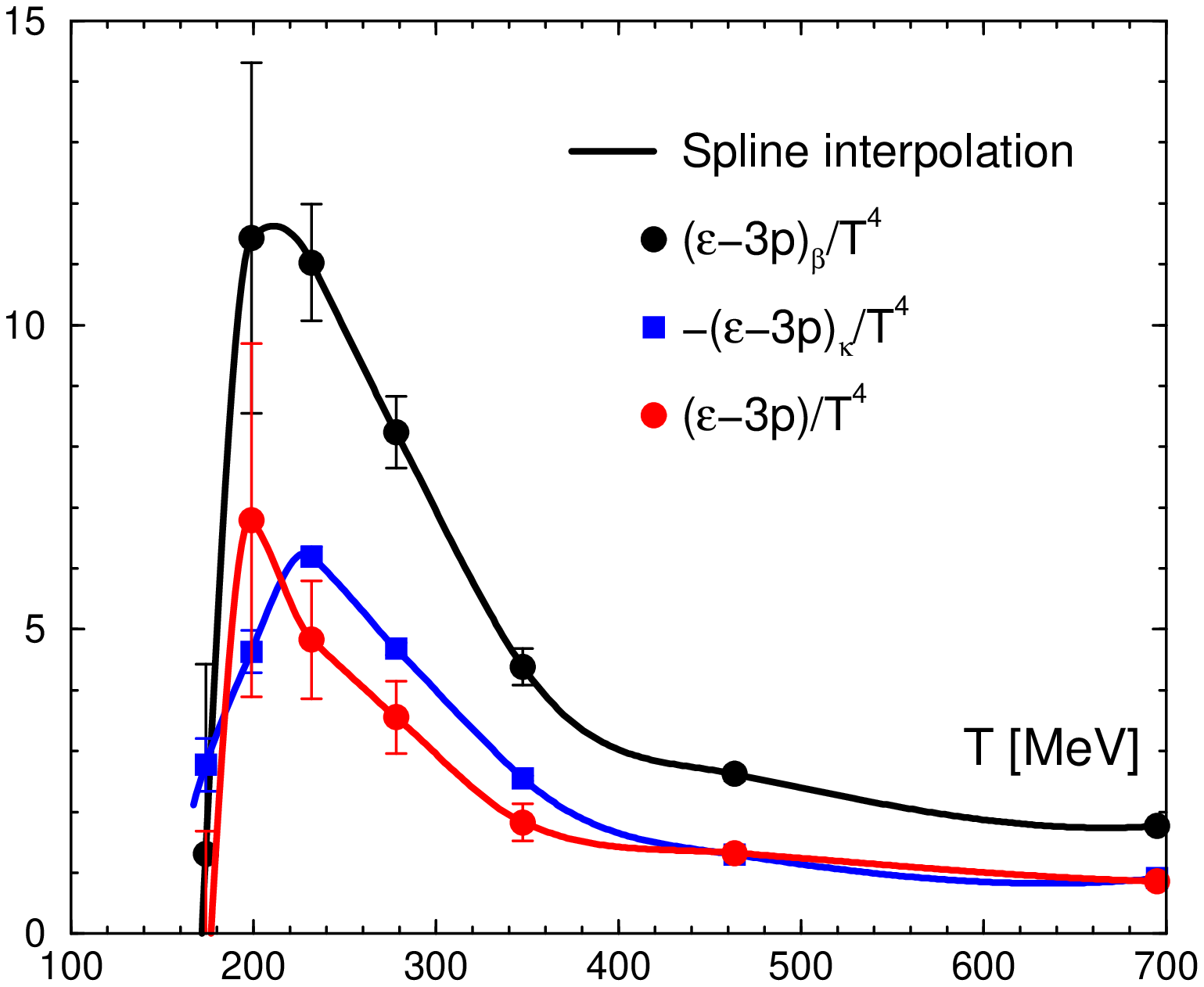} & 
    \includegraphics[width=70mm]{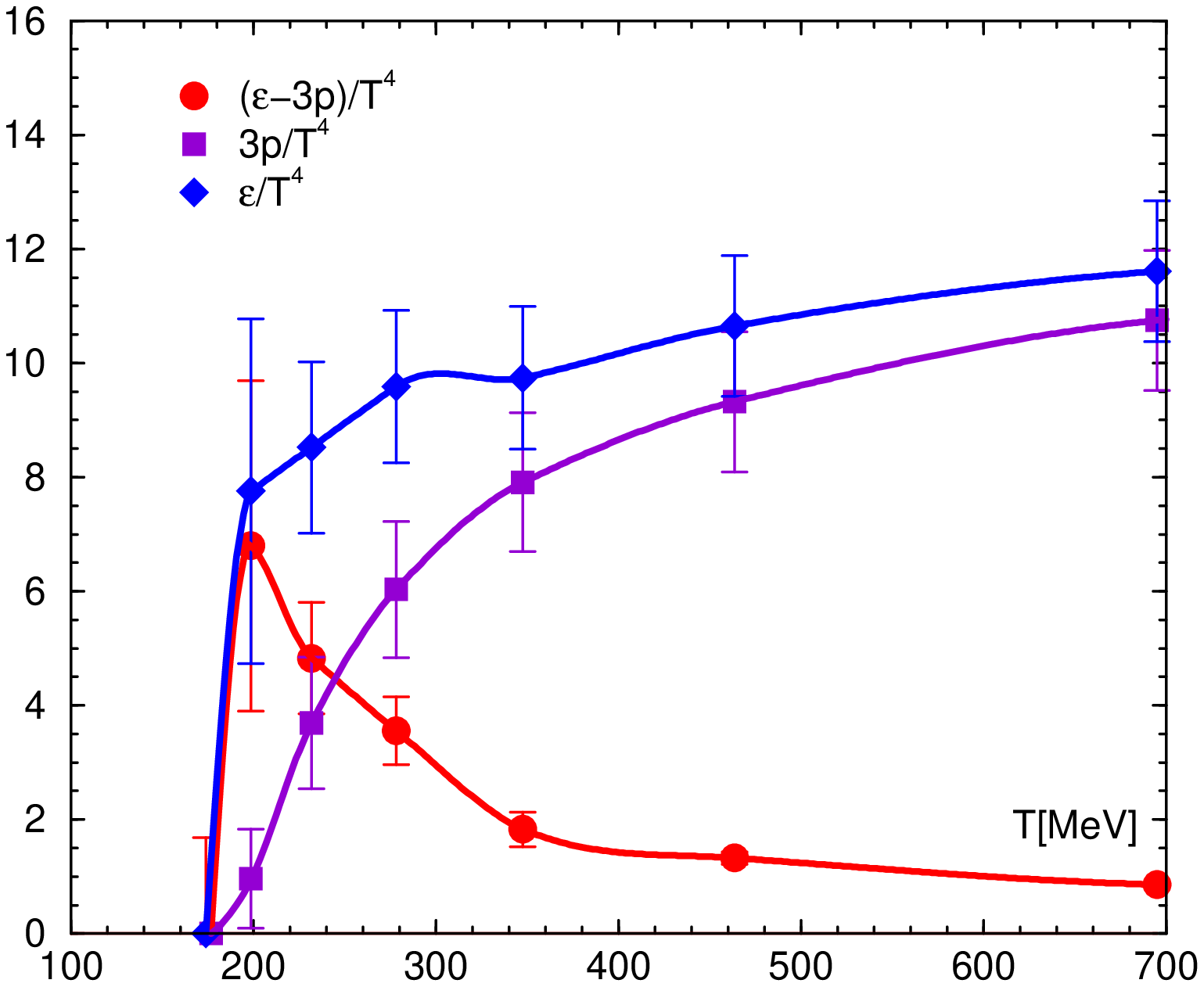}
    \end{tabular}
    \caption{
    {\em Left:} The trace anomaly and its $\beta$ and $\kappa$'s
    derivative contributions are shown. 
    The latter contribution is a minus
    quantity. The lines are drawn by a spline interpolation.
    {\em Right:} The trace anomaly and EOS, $\epsilon/T^4$
    and $3p/T^4$ are shown. 
    The lines are drawn by a spline interpolation.
    }
    \label{fig3}
  \end{center}
\end{figure}

The left panel of Fig.~\ref{fig3} shows the results for the trace anomaly, 
together with its decomposition into $\beta$ and $\kappa$ derivative parts in (\ref{eq:tranom}). 
The lines are spline interpolations.
We find that there is a big cancellation between the $\beta$ and $\kappa$ derivative parts and the resulting peak height of the trace anomaly is about 7.
This relatively low peak height obtained at $N_t \approx 14$ is roughly consistent with recent results from highly improved staggered quarks obtained on $N_t = 6$--12 lattices using the conventional fixed $N_t$ approach \cite{Borsanyi:2010cj,Bazavov:2010sb}.

Carrying out the $T$-integration (\ref{eq:Tintegral}) using a trapezoidal interpolation of the trace anomaly, 
we obtain the pressure $p/T^4$ shown in the right panel of Fig.~\ref{fig3}.
Here, we have chosen the starting point of the integration to be at $N_t=16$ where the trace 
anomaly vanishes within the statistical error.
The energy density $\epsilon/T^4$ is calculated by $p/T^4$ and $(\epsilon-3p)/T^4$.
The lines in the figure are spline interpolations.

The overall large errors in $p/T^4$ and $\epsilon/T^4$ are propagated from the large errors in $(\epsilon-3p)/T^4$ at low temperatures in the numerical integration.
From the left panel of Fig.~\ref{fig3}, we note that the large error in $(\epsilon-3p)/T^4$ at $T \sim 200$ MeV is due to the  $\beta$ derivative part in (\ref{eq:tranom}). 
We are currently trying to increase the statistics on $N_t=12$--16 lattices.

At the same time, we are starting simulations at $\beta=1.9$ to study the scaling.
Corresponding temperatures at $\beta=1.9$ are shown in the left panel of Fig.~\ref{fig1}. 
Smaller $N_t$ in the low temperature region will make the statistical error problem less severe.
Another objective to study at $\beta=1.9$ is to extend the investigation of EOS to lighter quark masses \cite{Aoki:2009ix}.

\section{Summary}
We presented the status of our project to calculate EOS for 2+1 flavor QCD with 
improved Wilson quarks.
Previous studies with Wilson-type quarks were limited to the case of two-flavor QCD with not quite light quarks \cite{CP-PACS-EOS}.
Adopting the fixed-scale approach, 
we succeed to calculate the first EOS for 2+1 flavor QCD with Wilson-type quarks on a fine lattice. 
Although the light quark masses are heavier than the physical values yet, 
our EOS looks roughly consistent with recent results with highly improved staggered 
quarks \cite{Borsanyi:2010cj,Bazavov:2010sb}.
On the other hand, we note that a large cancellation in the zero-temperature subtraction of the gauge action in the low temperature region leads to large statistical errors in the final EOS.
We need a large statistics there.
However, because the problem is limited on several lattices, we think that the overall computational cost is much smaller than that required with the conventional fixed $N_t$ approach, and tractable with current computer powers.
We are now ready to start a study with much lighter quarks, adopting the on-the-physical-point configurations by the PACS-CS Collaboration \cite{Aoki:2009ix}.

\section*{Acknowledgments}
We thank the members of the CP-PACS and JLQCD Collaborations for 
providing us with their high-statistics 2+1 flavor QCD configurations with improved Wilson quarks.
This work is in part supported 
by Grants-in-Aid of the Japanese Ministry
of Education, Culture, Sports, Science and Technology, 
( Nos.22740168, 
21340049  
22840020  
20340047  
) and the Grant-in-Aid for Scientific Research on Innovative Areas
(Nos.20105001, 20105003 
).  
This work is in part supported also by the Large Scale Simulation Program of High Energy Accelerator Research Organization (KEK) Nos. 09/10-25 and 10-09.

\end{document}